\documentclass[11pt]{article}
\usepackage[top=3cm,bottom=2.8cm,left=2.6cm,right=2.6cm]{geometry}

\usepackage{hyperref}
\hypersetup{colorlinks,citecolor=black,filecolor=black,linkcolor=black,
urlcolor=black,bookmarksnumbered}

\usepackage{amsmath,amsfonts,amssymb,amsthm,amscd,bbm,mathrsfs}
\usepackage{cancel}
\usepackage{listings}
\usepackage{moreverb}
\usepackage{booktabs}
\usepackage{epsfig}
\usepackage{cite}
\usepackage{algorithm}
\usepackage[]{algpseudocode}
\usepackage{authblk}

\providecommand{\keywords}[1]{\noindent{\textbf{\textit{Keywords---}}} #1}

\begin{document}
\date{}

\title{Dual Algorithms}

\author[a]{F. Pe\~nu\~nuri\thanks{{\it Corresponding Author. Tel:
 + 52 999 9300550x1051, 1052\\
E-mail addresses:\;francisco.pa@correo.uady.mx (F. Pe\~nu\~nuri),
osvaldo.carvente@correo.uady.mx (O. Carvente),
miguel.zambrano@correo.uady.mx (M. A. Zambrano-Arjona),
cacruz@cinvestav.mx (Carlos A. Cruz-Villar).
}}}
\author[a]{O. Carvente}
\author[a]{M. A. Zambrano-Arjona}
\author[b]{Carlos A. Cruz-Villar}
\affil[a]{Facultad de Ingenier\'ia, Universidad 
Aut\'onoma de Yucat\'an, A.P. 150, Cordemex, M\'erida, Yucat\'an, 
M\'exico.}
\affil[b]{Cinvestav-IPN, Departamento de Ingenier\'ia 
El\'ectrica, Av. IPN 2508, A. P. 14-740, 07300, M\'exico D.F., M\'exico.
}

%
%
%

\maketitle

\begin{abstract}
The cubic spline interpolation method, the Runge--Kutta method, and the 
Newton--Raphson method are extended to dual versions (developed in the 
context of dual numbers). This extension allows the calculation of the 
derivatives of complicated compositions of functions which are not 
necessarily defined by a closed form expression. The code for the
algorithms has been written in Fortran and some examples are presented. 
Among them, we use the dual Newton--Raphson method to obtain the 
derivatives of the output angle in the RRRCR spatial mechanism; we use 
the dual normal cubic spline interpolation algorithm to obtain the 
thermal diffusivity using photothermal techniques; and we use the dual 
Runge--Kutta method to obtain the derivatives of functions depending on 
the solution of the Duffing equation. 

\end{abstract}

\section{Introduction}
Analogous to a complex number $z = a\, + \, i\, b$ where $a$ and $b$ are 
real numbers and $i^2 = -1$, a dual number is defined as 
$\hat{r} = a\,+\,\epsilon\,b $ with $a$ and $b$ real numbers and 
$\epsilon^2 = 0$. Such numbers were introduced by Clifford who also 
developed their algebra in the late nineteenth century \cite{Clif1873}.

The fact that $\epsilon ^2 = 0$ suggests that the dual numbers can be 
used to differentiate functions, since in analogy to an infinitesimal 
$dx$, quantities of order $dx^n$ with $n$ an integer greater than or 
equal to two are usually neglected. It turns out that this reasoning 
is correct. This can be easily proved for polynomials and then via the 
Taylor Series, generalized for any analytic function. So, 
extending a real function to a dual function one can numerically obtain 
its derivatives. Nonetheless, most of the applications of dual 
numbers are in the area of mechanics (see for example 
\cite{Ettore2007} where  some contributions to mechanics based on dual 
numbers are presented) and only relatively recently have they been used 
to obtain the derivatives of functions \cite{Leuck1999, Piponi2004, 
Jefrey2011, Wenbin2013}.  Moreover, there are no papers addressing 
the \textit{dualization} (from now on the term 
\textit{dualization} will refer to extending a function or algorithm to 
the context of dual numbers) of algorithms such as the 
Newton--Raphson algorithm, the Runge--Kutta algorithm, or the cubic 
spline interpolation method. The present paper shows that a dualization 
of these algorithms allows the numerical calculation of the derivatives 
of complicated compositions of functions which frequently arise in 
science and engineering applications.
 
The problem of numerical differentiation has been addressed by many 
researchers. A complete review of the literature on this topic is 
beyond the scope of this paper. Nevertheless we want to cite some 
works which represent most of the techniques used to obtain 
derivatives numerically \cite{Liness1967, Rowlands1973, Griewank1989, 
Knowles1995, Khan1999, Khan2001, Li2005, Zhi2006, Karsten2007,Ashok2009,
Wang2010, Fang2010, Neidinger2010, Jefrey2011, Ramachandran2012, 
Dmitriev2012, Hassan2012, Wenbin2013}. It is worthwhile to mention that 
if there is a closed form expression of the function to be 
differentiated, the automatic differentiation methods (AD) 
\cite{Griewank1989, Neidinger2010, Wenbin2013} are (from the point of
view of precision, accuracy, and efficiency) the best choice to
calculate the derivatives. In the case when the derivatives of a set of
data are required, two excellent approaches are presented in 
\cite{Li2005,Hassan2012}.  However, in some applications, one deals with 
derivatives of the composition of functions defined by an expression 
rather than an explicit function or set of data. For example, one could 
be interested in calculating the derivatives of the composition of a 
function with another that is the solution of some differential 
equation, or in calculating the derivatives of a function which depends
on another function which is the spline interpolation of some data. In 
such cases, the use of dual numbers is especially suited for calculating
the derivatives. As illustrative examples, the thermal diffusivity for 
solids is obtained by making a dual cubic spline interpolation of the 
amplitude of the photothermal radiometry signal \cite{Depriestier2005}. 
The derivatives of functions depending on the solution of the Duffing 
equation \cite{Wiggins1990,Nijmeijer1995,Elcin2006} are calculated using 
the dual Runge--Kutta method. And the dual Newton--Raphson algorithm 
is used to obtain the derivatives of the output angle  in the RRRCR 
spatial mechanism \cite{Rico2014}. The Fortran code of the elemental 
dual functions as well as the dual version for the mentioned algorithms 
are provided as additional material to this article and can be 
downloaded from \url{http://dual-algorithms.frp707.esy.es/}. From these
codes, the dualization of many other functions and algorithms can be 
obtained.

\section{Derivatives using dual numbers}
How dual numbers can be used for calculating numerical 
derivatives can be seen in \cite{Gu1987,Cheng1994}. Here we 
briefly review the essential ideas, bearing in mind a numerical 
implementation.

Let $f:\mathbb{R}\to\mathbb{R}$ be an analytic function. Expanding this 
function in a Taylor Series and evaluating  at $\hat{x} = x + \epsilon$
(note that we have taken the dual part to be 1) we get
\begin{equation}
f(x+\epsilon) = f(x) + f'(x) \epsilon +\cancelto{0} {O(\epsilon^2)}.
\end{equation}
As we can see, $f(x) + f'(x) \epsilon$ is a dual number, so we associate 
the dual function
\begin{equation}\label{feps}
\hat{f}(\hat{x}) = f(x) + f'(x) \epsilon
\end{equation}
to it. As in the case of the complex numbers, there is an isomorphism 
between the dual numbers and $\mathbb{R}^2$, thus a dual number can be 
written as
\begin{equation}\label{r2nota}
\hat{r} = \{a,b\}.
\end{equation}
From now on we will use the notation given by Eq. (\ref{r2nota}), 
thus Eq. (\ref{feps}) will be written as

\begin{equation}\label{fd1x}
\hat{f}(\hat{x}) = \{f_0, f_1 \},
\end{equation}
where $f_0 = f(x)$ and $f_1 = f'(x)$.

The next step is to dualize the composition of $f(x)$ with another 
function $g(x)$. From the chain rule, we get
\begin{equation}\label{fdual1}
\hat{f}(\hat{g})  = \{f_0(g_0), f_1(g_0) g_1 \}.
\end{equation} 
For instance, if $h(x) = f(g(u(x)))$, the dual component of 
$\hat{h}(\hat{x})=\hat{f}(\hat{g}(\hat{u}(\hat{x})))$ will be $h'(x)$.
That is the power of the dual method of obtaining derivatives. We only 
need to implement the chain rule once. This makes the dual number method 
of obtaining derivatives an AD method.

The generalization to second derivatives is straightforward. To this end 
we define a 
\emph{new dual} number
\begin{align}\label{newdual}\nonumber
\tilde{r}&= a + b \, \epsilon_1 + c\,\epsilon_2, \\
\tilde{r}&=\{a,b,c\},
\end{align}
 with $a$, $b$ and $c$ being 
real numbers and $\epsilon_1$ and $\epsilon_2$ having the following 
multiplication table
\begin{equation}\label{TabMult}
\vbox{\tabskip0.5em\offinterlineskip
    \halign{\strut$#$\hfil\ \tabskip1em\vrule&&$#$\hfil\cr
    ~   & 1   &  \epsilon_1  & \epsilon_2  \cr
    \noalign{\hrule}\vrule height 12pt width 0pt
    1   & 1      &\epsilon_1   & \epsilon_2 \cr
    \epsilon_1 &\epsilon_1    & \epsilon_2  & 0     \cr
    \epsilon_2 &\epsilon_2    & 0  & 0 \cr
}}.
\end{equation}

Now, evaluating the Taylor expansion for $f(x)$ in 
$\tilde{x}=x+\epsilon_1$ we have,
\begin{align}\label{dual3fa}
f(\tilde{x}) &= f(x)+f'(x)\epsilon_1 + \frac{1}{2!}f''(x)\epsilon_1^2,\\
f(\tilde{x}) &= f(x) + f'(x)\epsilon_1 + \frac{1}{2!}f''(x)\epsilon_2. 
\label{dual3fb}
\end{align}
These expressions are of the form of Eq. (\ref{newdual})
so we can write
\begin{equation}\label{dual3f}
 \tilde{f}(\tilde{x}) = \{f(x),\; f'(x),\;  \frac{1}{2!}f''(x) \}.
\end{equation}

The fact that the third component is a half of the second derivative
does not matter. We can choose 
\begin{equation}
 \tilde{f}(\tilde{x}) = \{ f(x),\; f'(x),\; f''(x) \}
\end{equation}
as our definition for this \emph{new dual} function (in fact, by 
choosing $\epsilon_1^2=2 \epsilon_2$ the factor 1/2 disappears). Also 
the inappropriate name, \emph{new dual} function, can be changed to dual 
function; leaving to the context of the problem (number of components of
$f$) if we are talking about a  simple dual function or of an extended 
dual function of 3 components. Clearly Eq. (\ref{dual3f}) can be
generalized to obtain higher order derivatives. For instance, to obtain 
derivatives until order $n$, define $\epsilon_0 =1$, $\epsilon_1$, 
$\epsilon_2,\dots$, $\epsilon_n$ having the multiplication table
\begin{equation}\label{gentabmult}
\epsilon_i \cdot \epsilon_j =
\begin{cases}
0 & \text{if } i + j > n,
\vspace*{0.3cm}\\
\epsilon_{i+j} & \text{otherwise},
\end{cases}
\end{equation}
with $i,~j=0,1,\dots,n$.

It is also interesting to prove Eqs. (\ref{dual3fa}, \ref{dual3fb}, 
\ref{dual3f}) from a matrix algebra
point of view.  To this end we define the following matrices having the  
multiplication table (\ref{TabMult}),
\begin{equation}
\mathbbm{1}=\left(
\begin{array}{ccc}
 1 & 0 & 0\\
 0 & 1 & 0\\
 0 & 0 & 1
\end{array}
\right),~~
\boldsymbol{\epsilon_1}=\left(
\begin{array}{ccc}
 0 & 1 & 0\\
 0 & 0 & 1\\
 0 & 0 & 0
\end{array}
\right),~~
\boldsymbol{\epsilon_2}=\left(
\begin{array}{ccc}
 0 & 0 & 1\\
 0 & 0 & 0\\
 0 & 0 & 0
\end{array}
\right).
\end{equation}

Now, since
\begin{equation}
\mathbf{X} = x \, \mathbbm{1} + \boldsymbol{\epsilon_1}
\end{equation}
is in its Jordan canonical form, we will have \cite{Higham2008}
\begin{align}\label{Fmat1}
f(\mathbf{X}) &=  \left(
\begin{array}{ccc}
 f(x) & f'(x) & f''(x)/2!\\
 0 & f(x) & f'(x)\\
 0 & 0 & f(x)
\end{array}
\right),\\
\label{Fmat2}
f(\mathbf{X}) &= f(x)\, \mathbbm{1} + f'(x)\, \boldsymbol{\epsilon_1} + 
\frac{1}{2!}f''(x) \, \boldsymbol{\epsilon_2}.
\end{align}
Thus, the derivatives of $f$ are determined by calculating the matrix 
function $f(\mathbf{X})$. This matrix function  can 
be calculated by using the Cauchy integral formula for operator-value 
functions (see for example Sec. 16.8 of \cite{Hassani2000}). 

A generalization of Eqs. (\ref{Fmat1}, \ref{Fmat2}) to obtain 
higher order derivatives, can be done by calculating $f(\mathbf{X})$ 
with
\begin{equation}\label{Xn}
\mathbf{X} = x \, \mathbbm{1}_{(n+1)\times (n+1)} + 
\boldsymbol{\epsilon}_{1;(n+1)\times (n+1)},
\end{equation}
$\left[\boldsymbol{\epsilon}_{\alpha;(n+1)\times (n+1)}\right]_{i,j}=
\delta_{i,j-\alpha}$; being $\delta_{i,j}$ the Kronecker's delta 
function, and $\alpha = 1, 2, \dots, n$.

From Eq. (\ref{dual3f}), the analog of 
$\hat{x} = \{x,1 \}$ will be $\tilde{x} = \{x,1,0 \}$
and the analog of Eq. (\ref{fd1x}) will be
\begin{equation}
\tilde{f}(\tilde{x}) = \{f_0, f_1, f_2 \},
\end{equation}
where $f_0 = f(x)$, $f_1 = f'(x)$ and $f_2 = f''(x)$.
Similarly, we will have
 \begin{equation}\label{fdual2}
\tilde{f}(\tilde{g})  = \{f_0(g_0), f_1(g_0) g_1, f_2(g_0)g_1^2 + 
f_1(g_0)g_2 \},
\end{equation}
for the composition of two dual functions.

Eq. (\ref{fdual2}), is of central importance in dualizing a function or 
algorithm.  Notice that, unlike the finite difference methods, the use 
of Eq. (\ref{fdual2}) to obtain derivatives does not have the problems
of truncation or cancellation errors.


\subsection{Fortran implementation}
In order to facilitate the use of the dual number giving by Eq. 
(\ref{newdual}), it is convenient to introduce a derived data type. In 
the Fortran programming language, this can be done as:
\begin{verbatim}
type, public :: dual2
    real(8) :: f0, f1, f2
end type dual2
\end{verbatim}
Now, the dual number of Eq. (\ref{newdual})
can be written in Fortran as {\tt dual2(a,b,c)}. This derived data type
is defined in the module called {\tt dual2\_mod}. Also many Fortran 
functions and operators are overloaded to deal with this 
kind of number. Such a module is included in the additional material to
this article.

\section{Dualization of algorithms}

\subsection{Dual Newton--Raphson algorithm}
Let us suppose that we are given equation
\begin{equation}\label{FNR}
F\left(u(x),x\right)=0
\end{equation}
and that $u'(x_0)$ is required. In the case when a closed form 
expression for $u(x)$ can be obtained, the derivatives can be calculated 
by writing $u(x)$ in its dual form $\tilde u(\tilde x)$. If there is no 
closed form expression for $u(x)$, applying the chain rule yields
\begin{equation}\label{FNRCR}
u'(x_0) = -\frac{1}{\partial F/ \partial u (u(x_0),x_0)}
\frac{\partial F}{\partial x}(u(x_0),x_0).
\end{equation}
Assuming that there are closed form expressions for $\partial F / 
\partial x$ and $\partial F /\partial u$, the derivative $u'(x_0)$ can 
be calculated by solving numerically Eq. (\ref{FNR}) and substituting 
$u(x_0)$ into Eq. (\ref{FNRCR}). If  $\partial F / \partial x$ and 
$\partial F /\partial u$  cannot be obtained in closed form, 
then there is no conventional, easy, accurate, and precise 
way to calculate $u'(x_0)$. However if we code the dual version of some 
numerical solution method to solve Eq. (\ref{FNR}), then  we will obtain 
$u(x_0)$ and automatically its derivatives.  In the present paper, we 
dualize the Newton--Raphson method  \cite{Press1995,Cohen2010}. 

Let $f(q)$ be a real function of a real variable. The Newton--Raphson 
method allows finding a solution of the equation $f(q) = 0$. Starting to 
look for a solution in $q_0$, the algorithm determines an approximate 
solution as\footnote{The convergence of the Newton--Raphson method is 
not a trivial subject. In some cases, a small variation of $q_0$ causes 
a convergence/divergence of the method. For a study of the convergence 
of the method, we refer the reader to \cite{Susanto2009}.}
\begin{align}\label{realNRalgo}\nonumber
q_1 &= q_0 - \frac{f(q_0)}{f'(q_0)}\\\nonumber
&\vdots \\
q_{n+1} &= q_n - \frac{f(q_n)}{f'(q_n)}.
\end{align}

In order to obtain $u'(x_0)$ from Eq. (\ref{FNR}) we proceed as follows:
Write $\tilde{F}$ for the dual version of the function $F$. Define the
dual starting point $\tilde{u_0} = \{u_0, 0, 0\}$. Since we need
$F'(u)$, we define   $\tilde{x_0} = \{x_0, 0, 0\}$ and $\tilde{u} = 
\{u_0, 1, 0\}$, then from $\tilde{f_1} = \tilde{F}(\tilde{u},
\tilde{x_0})$ we obtain the first and second derivatives of $F$ with 
respect to $u$. This is an extra bonus of the method: we do not need to 
worry about the derivative of $F$. In practice, this is an issue, and 
the derivative must be provided by hand. Finally, write 
Eq. (\ref{realNRalgo}) in its dual form, keeping in mind that since we 
want $u'(x)$ we need to evaluate $\tilde{F}$ in $\tilde{u_0}$ and 
$\tilde{x}=\{x_0,1,0\}$. 

Putting  $\tilde{u_0}=$ \verb+u0d+,  $\tilde{x_0}=$ 
\verb+x0d+, and $\tilde{F}=$ \verb+fd+;  the 
essential part of the method in Fortran is

\begin{verbatim}
x0d = x0
u0d = u0
do k = 1, n
    fx  = fd(u0d, dual2(x0,1d0,0d0))
    fu  = fd(dual2(u0d%f0,1d0,0d0), x0d)
    u0d = u0d - fx/(fu%f1)
end do
\end{verbatim} 
This algorithm will produce $\tilde{u}(x)$. Using  Eq. (\ref{fdual2}), 
we can construct the general dual function for $u$:
$\tilde{u}(\tilde{g})$.
The complete algorithm is coded in  the module 
{\tt NR\_dual\_mod} of the additional material. We also coded Halley's 
method \cite{Gander1985}, so the user can choose between the simple 
Newton--Raphson method or Halley's method. Some examples are presented 
in Section \ref{secex}.

\subsection{Dual cubic spline interpolation} 
Let $P:A\subset \mathbb{R} \to B\subset \mathbb{R}$ 
be the function representing a cubic spline interpolation and let $f$ 
be a real function  such that  $f(P(x),x)$ is defined. The problem is to 
find  $ f'(P(x),x)$ with $x \in A$. We solve this problem by writing the 
dual version of the natural cubic spline interpolation.

\subsubsection*{Natural cubic splines}
Suppose that we have $n$ data points  $\left\lbrace (x_1,y_1),\dots,
(x_n,y_n)\right\rbrace$, $n>1$. A cubic spline is a spline constructed 
of piecewise third-order polynomials
\begin{equation}
 Y_i(t)=a_i\,+\,b_i\,t\,+\,c_i\,t^2\,+\,d_i\,t^3; ~~t\in[0,1],~~i=1,
 \dots,n-1,
\end{equation}
which pass through this set of points.
 
In determining the coefficient of the $i$th piece of the spline, a 
linear system of $4(n-1)-2$ equations and $4(n-1)$ unknowns will appear 
\cite{Bartels1987}. Two more equation can be obtained by demanding that 
$Y''(0)=0$ and $Y''(1)=0$. This yields the so called  natural cubic 
spline interpolation. With this, the coefficients of $Y_i$ are given by 
 \begin{align}  \nonumber
 a_i&=y_i\\ \nonumber
 b_i&=D_i\\  \nonumber
 c_i& = 3(y_{i + 1} - y_i) - 2 D_i - D_{i + 1}\\ 
  d_i&=2 (y_i - y_{i+ 1}) + D_i + D_{i + 1}
 \end{align}
where the  $D_s$  numbers are determined by solving the symmetric 
tridiagonal system:
\begin{equation}
\mathbf{T}\,\mathbf{D} \,=\,\mathbf{R}
\end{equation}
with
\begin{equation}\label{Tmatrix}
\mathbf{T}=\left(
\begin{array}{ccccccc}
 2 & 1 &  &  &  &  &  \\
 1 & 4 & 1 &  &  &  &  \\
  & 1 & 4 & 1 &  &  &  \\
  &  & \ddots & \ddots & \ddots &  &  \\
  &  &  & 1 & 4 & 1 &  \\
  &  &  &  & 1 & 4 & 1 \\
  &  &  &  &  & 1 & 2 \\
\end{array}
\right)
\end{equation}
\begin{equation}
\mathbf{D}=\left(
\begin{array}{c}
  D_1  \\
 D_2\\
 D_3\\
  \vdots \\
 D_{n-2} \\
  D_{n-1}\\
  D_n
\end{array}
\right)
\end{equation}
\begin{equation}
\mathbf{R} = 3\left(
\begin{array}{c}
 y_2-y_1 \\
 y_3-y_1  \\
 y_4-y_2\\
  \vdots \\
  y_{n-1}-y_{n-3}\\
 y_{n}-y_{n-2} \\
  y_{n}-y_{n-1}\\
\end{array}
\right).
\end{equation}
The inversion of an $n\times n$ tridiagonal matrix can be done by an 
$O(n)$ algorithm \cite{Press1995}, and in general it is this kind of 
algorithm which is used in order to find the inverse of the matrix 
$\mathbf{T}$ in Eq. (\ref{Tmatrix}). However, as we will show, an 
analytical formula for the inverse of $\mathbf{T}$ can be obtained.

Let us consider the $n\times n$ nonsingular tridiagonal matrix 
\[
\mathbf{M} = \left(
\begin{array}{ccccc}
 a_1 & b_1 &  &  &  \\
 c_1 &  a_2 & b_2 &  &  \\
  & c_2 &  \ddots & \ddots &  \\
  &  & \ddots & \ddots & b_{n-1} \\
  &  &  & c_{n-1} & a_n \\
\end{array}
\right).
\]

The inverse of such a matrix can be written as \cite{Usmani1994}:
\begin{equation}\label{Minverse}
\mathbf{M}^{-1}_{ij} =
\begin{cases}
(-1)^{i+j}\,b_i\dots b_{j-1}\,\theta_{i-1}\,\phi_{j+1}/\theta_n & 
\text{if } i \leqslant j,
\vspace*{0.3cm}\\
(-1)^{i+j}\,c_j\dots c_{i-1}\,\theta_{j-1}\,\phi_{i+1}/\theta_n & 
\text{if } i > j,
\end{cases}
\end{equation}
where $\theta_i$ is obtained by solving the recurrence equation
\begin{equation}\label{thetai}
\theta_i=a_i\,\theta_{i-1}\, - \, b_{i-1} c_{i-1} \theta_{i-2}, 
\text{ for } i=2,\dots , n,
\end{equation}
with $\theta_0 = 1$ and $\theta_1=a_1$. Then $\phi_i$ is obtained by 
solving the recurrence equation
\begin{equation}\label{phii}
\phi_i=a_i\,\phi_{i+1}\, - \, b_{i} c_{i} \phi_{i+2}, \text{ for } 
i=n-1,\dots , 1,
\end{equation}
with $\phi_{n+1}=1$ and $\phi_n = a_n$.

The use of Eq. (\ref{Minverse})  is not an efficient way to 
find the inverse of a tridiagonal matrix. However it can be used to 
deduce an analytical formula for the inverse of the matrix $\mathbf{T}$. 

Applying Eqs. (\ref{Minverse}, \ref{thetai} and \ref{phii}) to 
Eq. (\ref{Tmatrix}), we get

\begin{equation}\label{thetaiparticular}
\theta_s =
\begin{cases}
\frac{\left(2-\sqrt{3}\right)^s+\left(2+\sqrt{3}\right)^s}{2} & 
\text{if } s \neq n,\vspace*{0.3cm}\\
\frac{\left(2+ \sqrt{3}\right)^n \left(-3 + 2 \sqrt{3}\right) \,-\, 
\left(2-\sqrt{3}\right)^n \left(3 + 2 \sqrt{3}\right)}{2} & 
\text{if } s = n, 
\end{cases}
\end{equation}

\begin{equation}\label{phiiparticular}
\phi_s =
\begin{cases}
\frac{\left(2-\sqrt{3}\right)^{1+n}\,\left(2+\sqrt{3}\right)^s\,+\,
(2-\sqrt{3})^s \,(2+\sqrt{3})^{1+n}}{2} & \text{if } s \neq 1,
\vspace*{0.3cm}\\
\frac{\left(2+ \sqrt{3}\right)^n \left(-3 + 2 \sqrt{3}\right) - 
\left(2-\sqrt{3}\right)^n \left(3 + 2 \sqrt{3}\right)}{2} & 
\text{if } s = 1 .
\end{cases}
\end{equation}

Defining $inv$ as
\begin{equation}
inv(s,k) = (-1)^{s + k}  \theta_{s - 1}\,\phi_{k + 1}/\theta_n ,
\end{equation}
writing explicitly $\phi_{k+1}/\theta_n$ and carrying out some 
elementary algebra, we obtain
\begin{equation}\label{IT2}
inv(s,k) = \frac{(-1)^{s+k}}{2}\frac{1+(\alpha^{\dagger}/\alpha)^{s-1}}
{\beta^{\dagger}-(\alpha^{\dagger}/\alpha)^n\,\beta}\left[\alpha\,
\mathrm{e}^{(k+1)\ln\alpha^{\dagger}\, +\, (s-1)\ln\alpha }\,+\, 
\alpha^{\dagger}\,\mathrm{e}^{(s+k)\ln \alpha \, + \, n\,
\ln(\alpha^{\dagger}/\alpha)} \right],
\end{equation}
with 
\begin{align}\label{aatbbt}\nonumber
\alpha &= 2\,+\,\sqrt{3}\\ \nonumber
\alpha^{\dagger} &= 2\,-\,\sqrt{3}\\ \nonumber
\beta &= 2\,\sqrt{3} + 3\\ 
\beta^{\dagger} &= 2\,\sqrt{3} - 3.
\end{align}
From this, the in verse of the matrix $\mathbf{T}$ is

\begin{equation}\label{IT1}
\mathbf{T}^{-1}_{sk}= 
\begin{cases}
 inv(s,k)~~ \text{if}~~ s \leqslant k,\\
 inv(k,s)~~ \text{if}~~ s > k .
\end{cases}
\end{equation}

Thus the coefficients $D_s$ are given by
\begin{equation}
D_s=\sum_{k=1}^n \mathbf{T}^{-1}_{sk}\mathbf{R}_k, ~s=1,\dots, n.
\end{equation}

Now that all the coefficients have been determined, the parametric 
equation for the interpolated points will be
\begin{equation}\label{paraminterp}
\mathbf{r}_i(t) = \left\lbrace x_i + (x_{i + 1} - x_i)\, t,~ Y_i(t) 
\right\rbrace.
\end{equation}
Eliminating the parameter $t$, the equation for the  $i$th polynomial 
of the normal cubic spline will be
\begin{equation}\label{explicitinterp}
P_i(x)=Y_i\left(\frac{x - x_i}{x_{i + 1} - x_i}\right), 
~x \in [x_i,x_{i+1}].
\end{equation}
The interpolated points in the whole interval $[x_1, x_n]$ are
\begin{equation}
P(x) = \bigcup_{i=1}^{n-1} P_i(x).
\end{equation}

Once $P(x)$ has been promoted to a dual function (see the module 
{\tt cubic\_spline\_dual\_mod} of the aditional material), the 
derivatives of an arbitrary function $f=f(P(x),x)$ for any 
$x \in [x_1,x_n]$\footnote{Thus, dualizing an interpolation method we 
can calculate the derivatives of experimental data. Nevertheless we 
recommend using the methodology presented in \cite{Li2005,Hassan2012} 
instead, unless the derivatives between the nodes $[x_i,x_{i+1}]$ are 
required.}, as well as $P'(f(x),x)$, are calculated by writing $f$ in 
its dual form. Some examples are presented in Section \ref{secex}.

\subsection{Dual Runge--Kutta algorithm}\label{RKSec}
Let us consider the following  ordinary differential equation (ODE)
\begin{equation}\label{ODE2}
f''(t) = F(t,f,f'),
\end{equation}
 with initial conditions $f(t_0)=f_0$ and $f'(t_0)=v_0$.
 Let $g(t)$ be a function defined in such a way that $f(g(t))$ and 
 $g(f(t))$ are defined. The problem we want to address is to find the 
 derivatives of the composition of the functions $f$ 
and $g$. 

\subsubsection*{Dual Runge--Kutta 4th order method}
One of the most often used methods for numerically solving ODEs is the 
Runge--Kutta Method \cite{Press1995, Cohen2010, Griffiths2010}. 
In order to calculate $f'(g(t))$, $f''(g(t))$, $g'(f(t))$ and  
$g''(f(t))$, we will dualize the  4th order Runge--Kutta method (RK4).
Putting $x_2 = f'$,  $x_1 = f$, $u_1(t,x_1,x_2) = x_2$, $u_2(t,x_1,x_2)= 
F(t,x_1,x_2)$, the RK4 method  produces $x_2$ and $x_1$ and hence
 $f''(t)$. From this,
the dual version of $f$ as a function of the real variable $t$ 
(see the {\tt rk4} function in the  module 
{\tt runge\_kutta\_dual\_mod} of the additional material) is
\begin{equation}
\tilde{f}(t)=\left\lbrace f(t), f'(t), f''(t) \right\rbrace.
\end{equation}

The dual version of $f$ for any dual variable $\tilde{u}$, namely
$\tilde{f}(\tilde{u})$, can be constructed following Eq. (\ref{fdual2}). 
(See the  {\tt rk4dual} function in module {\tt runge\_kutta\_dual\_mod} 
of the additional material). 
Once we have $\tilde{f}(\tilde{u})$, the derivatives are calculated from 
the compositions $\tilde{f}(\tilde{g}(\tilde{t}))$
and $\tilde{g}(\tilde{f}(\tilde{t}))$---actually, we can calculate 
the derivatives of any function $G(f(t),t)$. Some examples are presented 
in Section \ref{secex}.

\section{Worked examples}\label{secex}
All the used functions are coded in the additional material to this 
article.

\subsection{Dual Newton--Raphson example 1}\label{DNRE1}
Suppose that $u(x)$ is defined by the equation $f(u,x)=0$ where
$$
f(u,x) = \cos(u\,x)-u^3+x+ \sin(u^2\,x).
$$
Let us consider the functions $g_1(x) = \sin(u(x)) + x$ and 
$g_2(x) = u(\sin x + x^2)$. The problem is to find $u(x_0)$, 
$u'(x_0)$, $u''(x_0)$, $g_1(x_0)$, $g_1'(x_0)$, $g_1''(x_0)$, 
$g_2(x_0)$, $g_2'(x_0)$, $g_2''(x_0)$, with $x_0=0.7$. This can be 
accomplished by writing the Newton--Raphson algorithm in its dual form. 
Such an algorithm is coded in the function 
\verb+NR_dual(NRkind,u0,n,fd,gdual)+.
In this function, \verb+NRkind+ is the kind of method used. 
\verb+NRkind = NR1+ is for the simple Newton--Raphson method and  
\verb+NRkind = NR2+  is for Halley's method,  \verb+u0+ is the dual 
point where the method starts to look for a solution, \verb+n+ is the 
number of iterations, \verb+fd+ is the equation to solve; in this 
example it will be $\tilde{f}(\tilde{u},\tilde{x})$, the dual version of 
$f(u,x)$; \verb+gdual+ is the dual point where we want to evaluate the 
functions. Writing the module \verb+equation_mod.f90+, where the 
function $f(u,x)$ is coded in its dual version, we have
\begin{listing}{1}
module equation_mod
    use dual2_mod
    implicit none
    
    contains

    function f(u,x) result(f_result)
    type(dual2), intent(in) :: u, x
    type(dual2)             :: f_result
    
    f_result = cos(u*x) - u**3 + x + sin(u**2*x)
    end function f
end module equation_mod 
\end{listing}

---a driver program could be as:

\begin{listing}{1}
include "dual2_mod.f90"
include "equation_mod.f90"
include "NR_dual_mod.f90"

program nr1
    
    use dual2_mod
    use NR_dual_mod
    use equation_mod
    
    implicit none
   
    real(8), parameter :: x = 0.7d0, u0 = 1.6d0
    type(dual2)        :: xd
    integer            :: n
    
    xd = dual2(x,1.d0,0.d0)
    
    n = 10
    
    print*,"The function u and its derivatives in x = 0.7 are"
    print*,NR_dual("NR1", u0, n, f, xd)
    print*," "
    print*,"The function g1 and its derivatives in x = 0.7 are"
    print*, sin(NR_dual("NR1", u0, n, f, xd)) + xd
    print*," "
    print*,"The function g2 and its derivatives in x = 0.7 are"
    print*,NR_dual("NR1", u0, n, f, sin(xd) + xd*xd)
    
end program nr1
\end{listing}
Notice the advantages of this method: we can calculate very complicated 
derivatives involving $u(x)$ by using the dual functions.
The results are shown in Table \ref{NREx}. 

\begin{table}[htb]
\caption{\label{NREx}Results for the Newton--Raphson example 1.}
\centering
\scalebox{0.9}{
\begin{tabular}{c c c c c c c c c}
\toprule $u(x_0)$ & $u'(x_0)$ & $''u(x_0)$ & $g_1(x_0)$ & $g_1'(x_0)$ &
$g_1''(x_0)$ &$g_2(x_0)$ & $g_2'(x_0)$ &$g_2''(x_0)$\\
\midrule 
1.3085 & 0.1163 & $-0.9337$ & 1.6658 & 1.0301 & $-0.2551$ & 1.2963 & 
$-0.2556$ & $-1.1425$\\
\bottomrule
\end{tabular}
}
\end{table} 

\subsection{Dual Newton--Raphson example 2}\label{DNRE2}
The example here presented concerns the RRRCR spatial mechanism 
\cite{Rico2014}. Following Eq. (2) of \cite{Rico2014} and the 
definitions given in the aforementioned reference, the output angle 
$\phi$ as a function of the 
input angle $\theta$ is given by the equation
\begin{align}\label{phitheta} \nonumber 
&a^2 c_1^2 c_2^2-2 a c_1 c_2^2 s_1 (b-e)-2 a
   c_1^2 c_2^2 L \cos \theta+2 a c_1
   c_2^2 R \cos \phi-c_1^2 c_2^2 (b-e)^2+\\ \nonumber
   &2c_1 c_2^2 L s_1 (b-e) \cos \theta+2c_1 c_2 L s_2 (b-e)\sin \theta-2
   c_1 R s_2 (b-e) \sin \phi-\\ \nonumber
   &2 R s_1 (b-e)\cos \phi+(b-e)^2-c_1^2 c_2^2 l^2+c_1^2
   c_2^2 L^2+c_1^2 c_2^2 R^2 \cos ^2\phi-\\ \nonumber
   &2c_1^2 c_2 L R \sin \theta \sin \phi-c_1^2 R^2 
   \left(1-2 \sin ^2\phi\right)-
   2c_1 c_2^2 L R \cos \theta \cos \phi-\\ 
   &2 c_1 c_2 L R s_1 s_2 \sin \theta \cos \phi+2 c_1 R^2 s_1 s_2 \sin
   \phi \cos \phi+R^2 \cos ^2\phi=0.
\end{align}

\begin{table}[htb]
\caption{\label{TparMech} Parameters used for the Newton--Raphson 
example 2.}
\centering
\scalebox{0.9}{
\begin{tabular}{c c}
\toprule Parameter & Value\\
\midrule $L$ & 0.3933578023  \\
         $l$ & 0.4174323687  \\
         $a$ & 0.9526245468 \\
         $R$ & 0.4484604992 \\
        $s_1$& 0.6298138891\\
        $s_2$& −0.2506389576\\
         $b$ & 2.0\\
         $e$ & 1.0 \\
\bottomrule
\end{tabular}
}
\end{table} 

Now, if one is interested in calculating the velocities and
accelerations, complicated functions involving $\phi$ and 
$\phi'(\theta)$ will appear. As an example, let us consider
the function $f(x) = 2\,\sin^2 x$ and the point $x_0=2.0$.
The problem is to calculate $f(\phi(x_0))$, $f'(\phi(x_0))$, 
$f''(\phi(x_0))$, $\phi(f(x_0))$, $\phi'(f(x_0))$ and $\phi''(f(x_0))$ 
for the same set of parameters given in \cite{Rico2014}, and reproduced 
in Table \ref{TparMech} for clarity.

Analogously to the  example in Section \ref{DNRE1}, we obtain the 
results shown in 
Table \ref{DerRes}. Notice that the solution of Eq. (\ref{phitheta}) 
for $\theta = 2$ is not unique. The values in Table  \ref{DerRes} 
are for $\phi = 20.9846$.

\begin{table}[htb]
\caption{\label{DerRes} Results for the Newton--Raphson example 2.}
\centering
\scalebox{0.9}{
\begin{tabular}{c c c c c c}
\toprule $f(\phi(x_0))$ & $f'(\phi(x_0))$ &$f''(\phi(x_0))$& 
$\phi(f(x_0))$ & $\phi'(f(x_0))$ &$\phi''(f(x_0))$\\
\midrule 
1.4279 & -1.7693 & -1.2856 &1.7817 &-1.6171  & -3.5137\\
\bottomrule
\end{tabular}
}
\end{table}

\subsection{Dual cubic spline interpolation example 1}\label{DCSIE1}
Let $y(x)$ be the normal cubic spline interpolation for the data shown 
in Table \ref{ptspline1}.
\begin{table}[htb]
\caption{\label{ptspline1} Points used for the dual cubic spline 
interpolation example 1.}
\centering
\scalebox{0.9}{
\begin{tabular}{l l}
\toprule $x$ & $y$ \\
\midrule 1. & 0 \\
 1.25 & 0.22314355 \\
 1.5 & 0.40546511 \\
 1.75 & 0.55961579 \\
 2. & 0.69314718 \\
 2.25 & 0.81093022 \\
 2.5 & 0.91629073 \\
 2.75 & 1.0116009 \\
 3. & 1.0986123 \\
 \bottomrule
\end{tabular}
}
\end{table} 
The values $y(x_0)$, $y'(x_0)$, $f(x_0)$, $f'(x_0)$, $g(x_0)$ 
and $g'(x_0)$ with $f(x) = x\,\sin^2 (y(x))$, $g(x)=y(x \sin^2 x)$ and 
$x_0=1.75$ can be calculated by dualizing the normal cubic spline 
interpolation method. This has been done in the function 
\verb+NCSplinedual(A,xd)+. In such a function, \verb+A+ is the matrix 
containing the points for which the spline will be constructed (in this
case, they would be the points of Table \ref{ptspline1}) and \verb+xd+
is the dual point where we want to evaluate, in this case 
\verb+xd=[1.75_dp,1_dp,0._dp]+. For instance the first (second) 
component of \verb+NCSplinedual(A,xd)+ will be $y(x_0)$ ($y'(x_0)$).

The derivatives $f'(x_0)$ and $g'(x_0)$, respectively, are obtained by 
taking component {\tt f1} of
\begin{verbatim}
xd*sin(NCSplinedual(A,xd))**2
\end{verbatim}
and
\begin{verbatim}
NCSplinedual(A, xd*sin(xd)**2)
\end{verbatim}
The results are given in Table \ref{csi1}.

\begin{table}[htb]
\caption{\label{csi1} Results for the dual cubic spline interpolation 
example 1.}
\centering
\scalebox{0.9}{
\begin{tabular}{c c c c c c}
\toprule $y(x_0)$ & $y'(x_0)$ & $f(x_0)$& $f'(x_0)$&$g(x_0)$&$g'(x_0)$\\
\midrule 
0.5596 & 0.5727 & 0.4931 & 1.1836 & 0.5272  & 0.2097\\
\bottomrule
\end{tabular}
}
\end{table} 

\subsection{Dual cubic spline interpolation example 2}\label{DCSIE2}
This example concerns the determination of the thermal properties of a 
solid using photothermal radiometry \cite{Depriestier2005}. In 
particular, we are interested in determining the thermal diffusivity
from the experimental data of Fig. 4 from \cite{Depriestier2005}. The 
experimental data can be extracted using EasyNData \cite{Uwer07}. 
Assuming that such data are stored in the matrix \verb+pts+, the dual 
cubic spline interpolation is done by
\verb+NCSplinedual(pts,x)+. Now, according to \cite{Depriestier2005}, 
the thermal diffusivity can be calculated from
\begin{equation}
\alpha_s=\frac{64\,L_s^2\,f_q}{9\,\pi},
\end{equation}
where $L_s = 522\,\mu$m is the thickness of the studied sample and $f_q$ 
is the frequency for which the derivative of the amplitude of the 
radiometry signal is zero. The frequency $f_q$ can be calculated using 
the Newton--Raphson algorithm. To this end we define the dual function 
\verb+fej2(x)+ containing the first and second derivatives of the 
spline. The essential part of the code in Fortran is
\begin{verbatim}
function fej2(x) result(f_result)
...
auxresult = NCSplinedual(pts,x)
f_result  = dual2(auxresult%f1,auxresult%f2,0.d0)
...
\end{verbatim}
 Taking \verb+x0 = 10.d0+ as the initial point where the Newton--Raphson 
 algorithm will start to look for a solution, $f_q$ can be found by

\begin{verbatim}
  ...
  fq = x0 
    
  do k=1,50
    fauxd = fej2(dual2(fq,1.d0,0.d0))
    fq    = fq - fauxd%f0/fauxd%f1
  end do
  ...
\end{verbatim}
 After this, the obtained thermal diffusivity was $\alpha_s=6.00
 \times10^{-6}$\,m\,$^2$s$^{-1}$ which is in good agreement with  the 
 value reported in \cite{Edwards69,Depriestier2005}.

\subsection{Dual Runge--Kutta example}\label{DRKE}
Consider the Duffing equation studied in \cite{Elcin2006}
\begin{equation}
f''(t)\, +\, 0.4\, f'(t)\,+\,1.1\,f(t)\,+\,f^3(t)\,=\,2.1\,\cos(1.8\,t)
\end{equation}
with initial conditions 
\begin{equation}
f(0) = 0.3, ~f'(0)= -2.3.
\end{equation}
The values $f(t)$, $(f\circ g)(t)$, $(g\circ f)(t)$, as well 
as their first and second derivatives at $t=1.0$ (or at some other $t$ 
where the functions are defined) for the function 
$g(t) = \sin t$ (or some other function where the above compositions are 
defined) can be calculated by dualizing the Runge--Kutta algorithm.
Such an algorithm is dualized in the function
\verb+rk4dual(u1,u2,t0,x10,x20,np,td)+. In this function, 
\verb+x10+ and   
\verb+x20+ are the initial conditions for $f(t_0)$ and $f'(t_0)$ 
respectively (see Section \ref{RKSec} for details and also for the 
definitions of $u_1$ and $u_2$); \verb+td+ is the dual point where we 
want to evaluate the solution;  and \verb+np+ is the number of steps 
between $t_0$ and and the real component of \verb+td+.   

In order to apply the function 
\verb+rk4dual(u1,u2,t0,x10,x20,np,td)+, it is convenient to 
write a module containing the functions $u_1(t,x_1,x_2)$ and 
$u_2(t,x_1,x_2)$. This can be done as follows:
\begin{listing}{1}
module equation_S45_mod
    implicit none
    contains

    function u1(t,x1,x2) result(f_result)
    real(8), intent(in) :: t, x1, x2
    real(8)             :: f_result
    
    f_result = x2 + 0d0*t + 0d0*x1
    
    end function u1
    
    function u2(t,x1,x2) result(f_result)
    real(8), intent(in) :: t, x1, x2
    real(8)             :: f_result
    
    f_result = 2.1d0*cos(1.8d0*t) - 0.4d0*x2 - 1.1d0*x1 - x1**3
    
    end function u2
end module equation_S45_mod
\end{listing}
From this, $f(t)$, $(f\circ g)(t)$,  $(g\circ f)(t)$, as well as their 
first and second derivatives at $t=1.0$ can be calculated with the 
following driver program:

\begin{listing}{1}
include "dual_modules/dual2_mod.f90"
include "dual_modules/runge_kutta_dual_mod.f90"
include "eqs_modules/equation_S45_mod.f90"

program test_S45_dual
    use dual2_mod
    use equation_S45_mod
    use runge_kutta_dual_mod

    implicit none
    
    integer, parameter :: np = 100
    real(8), parameter :: x10 = 0.3d0, x20 = -2.3d0, t0 = 0.d0
    type(dual2)        :: solution0, solution1, solution2
    type(dual2)        :: tvard
    
    tvard = dual2(1.d0,1.d0, 0.d0)
    
    solution0 = rk4dual(u1,u2,t0,x10,x20,np,tvard)
    solution1 = sin(rk4dual(u1,u2,t0,x10,x20,np,tvard))
    solution2 = rk4dual(u1,u2,t0,x10,x20,np,sin(tvard))
    
    print*,solution0
    print*,solution1
    print*,solution2
    
end program test_S45_dual
\end{listing} 
Assuming that this program is saved in the file \verb+test_S45_dual.f90+, 
we can compile it by executing the Fotran compiler in a terminal 
window\footnote{For instance, if the intel\textregistered~ Fortran 
compiler is used, {\tt ifort test\_S45\_dual.f90} will produce the 
executable file.}. The results are shown in Table \ref{RKex}.

\begin{table}[htb]
\caption{\label{RKex}Results for the Runge--Kutta example. The functions 
are evaluated at $t=1.0$.}
\centering
\scalebox{0.9}{
\begin{tabular}{c c c c c c c c c}
\toprule $f$ &$f'$&$f''$ &$(f\circ g)$ & $(f\circ g)'$ & $(f\circ g)''$& 
$(g\circ f)$ & $(g\circ f)'$ &$(g\circ f)''$ \\
\midrule 
$-0.7474$ & $-0.1282$ & 0.8140 & $-0.6797$ & $-0.0940$ & 0.6081 & 
$-0.7144$ & $-0.1638$ & 0.6608\\
\bottomrule
\end{tabular}
}
\end{table} 

\section{Conclusions}
After a dualization of the normal cubic spline interpolation algorithm,
the Runge--Kutta algorithm, and the Newton--Raphson algorithm, it is 
possible to calculate the derivatives of functions efficiently, 
precisely, and accurately. So we can calculate the derivatives of 
functions depending on the solution of algebraic  or differential 
equations as well as functions resulting from the spline interpolation
of experimental data.  As an added value to the normal cubic spline 
interpolation algorithm, it is shown that a closed form expression for 
its coefficients can be obtained. Interesting applications in science 
and engineering were studied. Those examples can be used as a guide to 
dualize many other functions and algorithms. For example, it would be 
interesting to dualize the trapezium rule although this would be only 
for academic purposes since there is not much to gain because its  
components would be the integral, the first derivative (which is 
actually the function to integrate), and the second derivative (which is 
actually the first derivative of the function to integrate). 
Nevertheless, the dualization of an integration method could be 
necessary for dualizing some other algorithm.

%


\begin{thebibliography}{100}
\bibitem{Clif1873}
W.~Clifford, Preliminary sketch of biquaternions, Proc. London Mathematical
  Society 1~(1-4) (1873) 381--395.

\bibitem{Ettore2007}
E.~Pennestr\`i, R.~Stefanelli, Linear algebra and numerical algorithms using
  dual numbers, Multibody System Dynamics 18 (2007) 323--344.

\bibitem{Leuck1999}
H.~Leuck, H.-H. Nagel, Automatic differentiation facilitates of-integration
  into steering-angle-based road vehicle tracking, IEEE Computer Society
  Conference on Computer Vision and Pattern Recognition 2~(5) (1999) 2360.

\bibitem{Piponi2004}
D.~Piponi, Automatic differentiation, {C++} templates, and photogrammetry,
  Journal of Graphics, GPU, and Game Tools 9~(4) (2004) 41--55.

\bibitem{Jefrey2011}
J.~A. Fike, J.~J. Alonso (Eds.), The Development of Hyper-Dual Numbers for
  Exact Second-Derivative Calculations, Proceedings of the 49th AIAA Aerospace
  Sciences Meeting, Orlando FL, USA, 2011.

\bibitem{Wenbin2013}
W.~Yu, M.~Blair, {DNAD}, a simple tool for automatic differentiation of
  {Fortran} codes using dual numbers, Computer Physics Communications 184
  (2013) 1446--1452.

\bibitem{Liness1967}
J.~N. Lyness, C.~B. Moler, Numerical differentiation of analytic functions,
  SIAM Journal on Numerical Analysis 4 (1967) 202--210.

\bibitem{Rowlands1973}
R.~E. Rowlands, T.~Liber, I.~M. Daniel, P.~G. Rose, Higher-order numerical
  differentiation of experimental information, Experimental Mechanics 13 (1973)
  105--112.

\bibitem{Griewank1989}
A.~Griewank, On automatic differentiation, in Mathematical Programming: Recent
  Developments and Applications, M. Iri and K. Tanabe, eds., Kluwer, Dordrecht,
  The Netherlands, 1998.

\bibitem{Knowles1995}
I.~Knowles, R.~Wallace, New finite difference formulas for numerical
  differentiation, Numerische Mathematik 70 (1995) 91--110.

\bibitem{Khan1999}
I.~Khan, R.~Ohba, Closed-form expressions for the finite difference
  approximations of first and higher derivatives based on {Taylor} series,
  Journal of Computational and Applied Mathematics 107 (1999) 179--193.

\bibitem{Khan2001}
I.~Khan, R.~Ohba, New finite difference formulas for numerical differentiation,
  Journal of Computational and Applied Mathematics 126 (2001) 269--276.

\bibitem{Li2005}
J.~Li, General explicit difference formulas for numerical differentiation,
  Journal of Computational and Applied Mathematics 183 (2005) 29--52.

\bibitem{Zhi2006}
Z.~Qian, C.-L. Fu, X.-T. Xiong, T.~Wei, Fourier truncation method for high
  order numerical derivatives, Applied Mathematics and Computation 181 (2006)
  940--948.

\bibitem{Karsten2007}
K.~Ahnert, M.~Abel, Numerical differentiation of experimental data: Local
  versus global methods, Computer Physics Communications 177 (2007) 764--774.

\bibitem{Ashok2009}
A.~K. Singh, B.~S. Bhadauria, Finite difference formulae for unequal
  sub-intervals using {Lagrange’s} interpolation formula, International
  Journal of Mathematical Analysis 3 (2009) 815--827.

\bibitem{Wang2010}
Z.~Wang, R.~Wen, Numerical differentiation for high orders by an integration
  method, Journal of Computational and Applied Mathematics 234 (2010) 941--948.

\bibitem{Fang2010}
F.-F. Dou, C.-L. Fu, Y.-J. Ma, A wavelet-{Galerkin} method for high order
  numerical differentiation, Applied Mathematics and Computation 215 (2010)
  3702--3712.

\bibitem{Neidinger2010}
R.~D. Neidinger, Introduction to automatic differentiation and {MATLAB}
  object-oriented programming, SIAM Review 52~(3) (2010) 545--563.

\bibitem{Ramachandran2012}
M.~Ramachandran, Fast derivative computation using smooth {X-splines}, Applied
  Numerical Mathematics 62 (2012) 1654--1662.

\bibitem{Dmitriev2012}
V.~I. Dmitriev, Z.~G. Ingtem, Numerical differentiation using spline functions,
  Computational Mathematics and Modeling 23 (2012) 179--193.

\bibitem{Hassan2012}
H.~Hassan, A.~Mohamad, G.~Atteia, An algorithm for the finite difference
  approximation of derivatives with arbitrary degree and order of accuracy,
  Journal of Computational and Applied Mathematics 236 (2012) 2622--2631.

\bibitem{Depriestier2005}
M.~Depriester, P.~Hus, S.~Delenclos, A.~H. Sahraoui, New methodology for
  thermal parameter measurements in solids using photothermal radiometry,
  Review of Scientific Instruments 76 (2005) 074902--1--074902--6.

\bibitem{Wiggins1990}
S.~Wiggins, Introduction to Applied Nonlinear Dynamical Systems and Chaos,
  Springer-Verlag, New York, 1990.

\bibitem{Nijmeijer1995}
H.~Nijmeijer, H.~Berghuis, On {Lyapunov} control of the {Duffing} equation,
  IEEE Transactions on Circuits and Systems--I~(42) (1995) 473--477.

\bibitem{Elcin2006}
{El{\c{c}}in Yusufo{\v{g}}lu}, Numerical solution of {Duffing} equation by the
  {Laplace} decomposition algorithm, Applied Mathematics and Computation~(177)
  (2006) 572--580.

\bibitem{Rico2014}
J.~J. Cervantes-S\'anchez, J.~M. Rico-Mart\'inez, V.~H. {P\'erez-Mu\~noz},
  A.~Bitangilagy, Function generation with the {RRRCR} spatial linkage,
  Mechanism and Machine Theory 74 (2014) 58--81.

\bibitem{Gu1987}
Y.-L. Gu, J.~Y.~S. Luh, Dual-number transformation and its applications to
  robotics, IEEE Journal of Robotics and Automation 3~(6) (1987) 615--623.

\bibitem{Cheng1994}
H.~H. Cheng, Programming with dual numbers and its applications in mechanisms
  design, Engineering with Computers 10~(4) (1994) 212--229.

\bibitem{Higham2008}
N.~J. Higham, Functions of Matrices: {Theory} and Computation, Society for
  Industrial and Applied Mathematics, Philadelphia, PA, USA, 2008.

\bibitem{Hassani2000}
S.~Hassani, Mathematical Physics. A modern Introduction to its Foundations,
  Springer, New York, 2000.

\bibitem{Press1995}
W.~H. Press, S.~A. Teukolsky, W.~T. Vetterling, B.~P. Flannery, Numerical
  Recipes in Fortran 77 The Art of Scientific Computing, 2nd Edition, Vol.~1,
  Cambridge Univerity Press, 1995.

\bibitem{Cohen2010}
H.~Cohen, Numerical Approximation Methods, Springer-Verlag, New York, 2010.

\bibitem{Susanto2009}
H.~Susanto, N.~Karjanto, Newton’s method’s basins of attraction revisited,
  Applied Mathematics and Computation 215 (2009) 1084--1090.

\bibitem{Gander1985}
W.~Gander, On {Halley's} iteration method, The American Mathematical Monthly 92
  (1985) 131--134.

\bibitem{Bartels1987}
R.~H. Bartels, J.~C. Beatty, B.~A. Barsky, An Introduction to Splines for Use
  in Computer Graphics and Geometric Modeling, Morgan Kaufmann, Los Altos, CA,
  1987.

\bibitem{Usmani1994}
R.~A. Usmani, Inversion of a tridiagonal jacobi matrix, Linear Algebra and its
  Applications 212–213~(0) (1994) 413 -- 414.

\bibitem{Griffiths2010}
D.~F. Griffiths, D.~J. Higham, Numerical Method for Ordinary Differential
  Equations, Springer, London, 2010.

\bibitem{Uwer07}
P.~Uwer, Easyndata: A simple tool to extract numerical values from published
  plots, arXiv: 0710.2896v1 [physics.comp-ph].

\bibitem{Edwards69}
A.~L. Edwards, A compilation of thermal property data for computer
  heat-conduction calculations, UCRL-50589, University of California Lawrence
  Radiation Laboratory, 1969.

\end{thebibliography}
\end{document}